\documentclass[letterpaper]{article} 
\usepackage{aaai2026preprint} 
\usepackage{times}  
\usepackage{helvet}  
\usepackage{courier}  
\usepackage[hyphens]{url}  
\usepackage{graphicx} 
\usepackage{natbib}  
\usepackage{caption} 
\usepackage{algorithm}
\usepackage{algpseudocode}
\usepackage{amsmath}
\usepackage{amssymb}
\usepackage{subcaption}
\usepackage{array}
\usepackage{todonotes}
\usepackage{longtable}
\usepackage{tikz}
\usepackage{pgfplots}

\usepackage{amsthm}
\newtheorem{theorem}{Theorem}
\newtheorem{corollary}{Corollary}
\newtheorem{proposition}{Proposition}
\theoremstyle{remark}
\newtheorem*{remark}{Remark}

\DeclareMathOperator*{\argmax}{arg\,max}

\usepackage{newfloat}
\usepackage{listings}
\DeclareCaptionStyle{ruled}{labelfont=normalfont,labelsep=colon,strut=off} 
\floatstyle{ruled}
\newfloat{listing}{tb}{lst}{}
\floatname{listing}{Listing}
\title{A Pressure-Based Diffusion Model for Influence Maximization on Social Networks}
\author{
    Curt Stutsman\textsuperscript{\rm 1},
    Eliot W. Robson\textsuperscript{\rm 2}\footnote{Part of this work was completed while the author was a Ph.D. student at the University of Illinois Urbana‑Champaign.},
    Abhishek K. Umrawal\textsuperscript{\rm 1}
}
\affiliations{

    \textsuperscript{\rm 1}University of Illinois Urbana-Champaign, Urbana, Illinois 61801, USA\\
    \textsuperscript{\rm 2}Narmi Inc., New York City, New York 10016, USA\\

    curtiss8@illinois.edu, eliot.robson24@gmail.com, aumrawal@illinois.edu
}

\begin{document}

\maketitle
\begin{abstract} 
In many real-world scenarios, an individual's local social network carries significant influence over the opinions they form and subsequently propagate. In this paper, we propose a novel diffusion model — \textbf{the Pressure Threshold model (PT)} — for dynamically simulating the spread of influence through a social network. This model extends the popular Linear Threshold (LT) model by adjusting a node's outgoing influence in proportion to the influence it receives from its activated neighbors. We examine the Influence Maximization (IM) problem under this framework, which involves selecting seed nodes that yield maximal graph coverage after a diffusion process, and describe how the problem manifests under the PT model. Experiments on real-world networks, supported by enhancements to the open-source network-diffusion library CyNetDiff, reveal that greedy IM under PT can yield seed sets distinct from those under LT. Furthermore, the analyses show that densely connected networks amplify pressure effects far more strongly than sparse networks.
\end{abstract}


\section{Introduction}
\subsection{Motivation}
Diffusion processes within human populations aim to imitate how behaviors, information, diseases, and more propagate through social networks. Diffusion models provide the underlying mathematics that dictate how attributes spread from one node to the next. Early diffusion models provided critical insights into these processes, paving the way for numerous practical applications in epidemiology and the social sciences. With the advent of digital platforms and the increasing significance of viral marketing around the turn of the twenty-first century \cite{evans2010social}, diffusion models became crucial for solving the Influence Maximization (IM) problem, first formulated by \citet{Domingos}. The goal of the IM problem is to select a set of $k$ seed nodes in a network such that the expected spread of influence is maximized after a diffusion process. The choice of diffusion model significantly impacts how influence propagates, thus determining the effectiveness of seed node selection.

The influence maximization (IM) problem was popularized by \citet{kempe2003maximizing}, who proposed a greedy approximation algorithm for solving it. Building on threshold-based models of collective behavior \citep{granovetter1978threshold} and cascade models from viral marketing \citep{goldenberg2001talk}, they formalized the Linear Threshold (LT) and Independent Cascade (IC) models as graph-based diffusion processes that remain widely used in IM research. These models seek to not only mathematically define the propagation of influence but also accurately represent genuine human interactions in real-world communities. However, despite their utility, traditional models often fail to capture key interpersonal phenomena such as reinforcement within echo chambers, where repeated exposure amplifies beliefs and behaviors \cite{misinformation}. Empirical work shows that online diffusion is shaped by local reinforcement: false or novel items spread farther, faster, and deeper than truth on Twitter \cite{vosoughi_science}, users self-organize into homophilic clusters that bias transmission toward like-minded peers \cite{cinelli_pnas2021,brugnoli_srep2019}, and adoption probabilities increase with multiple confirming exposures, i.e., complex contagion \cite{monsted_plosone2017}.

To address these shortcomings and better represent realistic social dynamics, this research introduces the Pressure Threshold model (PT), an extension of the LT model. The PT model introduces dynamic feedback mechanisms, wherein nodes amplify their outgoing influence proportional to the incoming influence from activated neighbors. \textcolor{black}{This mechanism directly operationalizes the empirically observed reinforcement pattern: when a node activates under substantial `pressure,' it becomes a stronger advocate in subsequent steps, thereby capturing within-community amplification consistent with observed diffusion regularities \cite{vosoughi_science,cinelli_pnas2021}.} Such reinforcement reflects realistic social scenarios. For example, individuals are more likely to strongly advocate a product when many of their friends or peers have already adopted it \cite{Gunawan_Rahmania_Kenang_2023}. \textcolor{black}{By making this feedback explicit while retaining LT’s additive semantics, PT offers a simple and interpretable way to model reinforcement beyond static LT/IC assumptions, aligning the diffusion mechanism with documented online behaviors and providing new leverage for IM analyses.}

\subsection{Literature Review}
Diffusion models have a long history of use, extending beyond the IM problem. We now review the history of diffusion models, their role in Influence Maximization, and dynamic approaches related to the Pressure Threshold model.

\subsubsection{Early Diffusion Models:}
The study of diffusion models originated in epidemiology and sociology. Early foundational work by \citet{SIR} introduced the Susceptible-Infected-Recovered (SIR) epidemiological model, setting the stage for understanding contagion processes. In social networks, \citet{granovetter1978threshold}'s threshold model became foundational, conceptualizing adoption based on cumulative peer influence. Both the Linear Threshold (LT) and Independent Cascade (IC) models were later formalized by \citet{kempe2003maximizing}, and remain standard in IM research. \textcolor{black}{These models provide the basis for most algorithmic work on influence spread.}

\subsubsection{Influence Maximization:}
The IM problem was explicitly formulated by \citet{Domingos} in the context of viral marketing, aiming to maximize influence spread through optimal seed selection. \citet{kempe2003maximizing} advanced IM research by proving NP-hardness, submodularity of influence, and introducing a greedy approximation algorithm, sparking extensive follow-up work.

\subsubsection{Improvements to IM:}
Subsequent research refined IM algorithms for scalability and efficiency. \citet{leskovec2007cost} introduced CELF using lazy forward evaluations, reducing complexity while preserving guarantees, later improved as CELF++ by \citet{CELF++}. \citet{RIS} developed Reverse Influence Sampling (RIS) for near-optimal scalability. \citet{Degree_Discount} proposed the degree discount heuristic for faster, though less optimal, solutions. \textcolor{black}{Recent work explores community-aware heuristics \cite{umrawal2023leveraging,robson2025community} and learning-based approaches \cite{kumar2022influence}.} Beyond the original problem, variants include partial incentives \cite{chen2020scalable,UmrawalAQ2023,BhimarajuRVU2024}, online settings \cite{agarwal2022stochastic,nie2022explore,xu2026lofa}, fairness \cite{becker2022fairness,lin2023fair,nguyen2022fairnessbudget}, and deep learning methods \cite{kumar2022influence}. See \cite{li2023surveyml} for a detailed survey.

\subsubsection{Dynamic Diffusion Models:}
Unlike the static LT and IC models, dynamic models capture evolving network structure and influence weights. \citet{Zhuang} addressed edge dynamics over time, and \citet{Xie} proposed DynaDiffuse, combining continuous-time propagation with dynamic edge weights. \textcolor{black}{These approaches share similarities with PT, which introduces adaptive influence updates.}

\subsubsection{Predictive Comparison of Diffusion Models:}
Some studies compare diffusion models against real-world cascades. \citet{Kuo} evaluated predictive accuracy, finding IC superior for direct prediction tasks, while LT performed better in indirect comparisons. \citet{Ananthasubramaniam2024TheRO} analyzed Twitter hashtag propagation, showing that accuracy depends on network topology and content type. \citet{AralWalker2012} validated cumulative influence in product adoption, supporting threshold-based assumptions.

\subsubsection{Shortcomings of Existing Models:}
Despite progress, existing models often overlook reinforcement dynamics evident in online networks and misinformation cascades. \citet{misinformation} showed that ideological reinforcement and echo chambers amplify diffusion on Facebook. \citet{vosoughi_science} demonstrated that false news spreads faster and deeper than truth, driven by repeated exposure and amplification. \textcolor{black}{Such findings underscore the need for models that incorporate local reinforcement, a gap addressed by the PT model.}

\textcolor{black}{In summary, while prior work has advanced IM theory and algorithms, gaps remain in modeling dynamic feedback and empirically observed reinforcement. Our research bridges these gaps by introducing a diffusion model that explicitly accounts for local social reinforcement, supported by empirical evidence.}

\subsection{Contribution}
This paper makes four primary contributions:

\begin{enumerate}
    \item \textbf{Pressure Threshold (PT) Model.} We introduce the Pressure Threshold diffusion model, an extension of the Linear Threshold (LT) model that incorporates local reinforcement: when a node activates, it increases its outgoing influence in proportion to the cumulative incoming pressure that triggered its activation. \textcolor{black}{This mechanism captures empirically observed amplification within communities while preserving the additive semantics of LT.}

    \item \textbf{Theoretical Properties.} We prove that influence maximization under PT is NP-hard (by reduction from LT). \textcolor{black}{We also show by counterexample that PT is neither monotone nor submodular when the amplification parameter $\alpha > 0$, clarifying the scope of classical greedy-approximation guarantees.}

    \item \textbf{Implementation.} We enhance the CyNetDiff \cite{robson2024cynetdiff} Python library for accelerated simulations, leveraging Cython-based kernels to support large-scale Monte Carlo evaluation across diffusion models, including PT. These improvements enable practical experiments on real-world networks at scale.

    \item \textbf{Empirical Analysis.} Across real and synthetic networks, PT (i) can produces seed sets distinct from LT and (ii) often achieves larger final spreads for comparable budgets. Although submodularity does not hold in general, we observe approximate submodular behavior in large networks for small $\alpha$, under which greedy strategies remain effective. Analyses on synthetic topologies highlight regimes where PT’s reinforcement is especially beneficial (e.g., denser structures with more opportunities for amplification).
\end{enumerate}

\noindent \textcolor{black}{Collectively, these results indicate that explicitly modeling local reinforcement yields diffusion dynamics and seed-selection behavior that are not captured by static LT/IC formulations, while remaining computationally tractable for influence maximization workflows.}

\subsection{Organization}
The remainder of the paper is organized as follows. 
Section~\ref{sec:preliminaries} introduces preliminaries and formally defines the Pressure Threshold model, including the NP-hardness argument. 
Section~\ref{sec:model_properties} provides a small-scale walk-through of PT dynamics and presents our analysis of non-monotonicity and non-submodularity using counterexamples. 
Section~\ref{sec:experiments} describes our experimental setup, implementation details (CyNetDiff), datasets, and results on both real-world and synthetic networks. 
Section~\ref{sec:conclusion} concludes with implications, limitations, and avenues for future work.

\section{Preliminaries and Model Formulation}\label{sec:preliminaries}
In this section, we formally establish foundational concepts and clearly define the Pressure Threshold model (PT). We discuss existing diffusion models, formalize the proposed PT model, and analyze its computational complexity.

\subsection{Linear Threshold Model and Influence Maximization}
Diffusion models mathematically describe how information, ideas, behaviors, or diseases propagate through a network. One of the most widely studied models in the context of influence maximization is the Linear Threshold (LT) model, which models adoption via cumulative social influence and classic threshold-based theories \cite{granovetter1978threshold}.

Given that the PT model is a generalization of the LT model, we will now give a formal definition of the LT model. Let $G = (V,E)$ be a directed graph where $V$ is the set of nodes and $E \subseteq V \times V$ is the set of edges. Each edge $(u, v) \in E$ is assigned a weight $w_{uv} \in [0,1]$ such that for each node $v$, $\sum_{u \in N(v)} w_{uv} \leq 1$, where $N(v)$ denotes the in-neighbors of $v$. The edge weight is meant to quantify the degree to which node \( u \) can influence node \( v \). Additionally, each node $v$ has a threshold $\theta_v$ drawn uniformly at random from the interval $(0,1]$. Initially, a set of seed nodes $S \subseteq V$ is activated. In each subsequent discrete time step, an inactive node $v$ becomes active if the total weight of its active in-neighbors meets or exceeds its threshold:
\[
\sum_{u \in P(v)} w_{uv} \geq \theta_v,
\]
where \( P(v) = \{ u \in A_t : (u, v) \in E \} \) denotes the set of active parent nodes (or in-neighbors) of \( v \). This process continues until no new nodes can be activated, and the diffusion ceases. \textcolor{black}{For clarity, we consistently use $N(v)$ for the (time-invariant) set of in-neighbors of $v$, and $P(v)$ for the (time-dependent) subset of $N(v)$ that are active at time $t$.}

We define the influence of a seed set $S \subseteq V$ under the Linear Threshold (LT) model as the expected number of nodes activated by the diffusion process initiated from $S$, where the expectation is taken over the random thresholds $\theta_v \in (0,1]$ for all $v \in V$. 
More generally, for a diffusion model $M$, we denote the influence of $S$ by $\sigma_M(S)$. The influence maximization (IM) problem is then to select, given a directed graph $G=(V,E)$, a diffusion model $M$, and an integer budget $k$, a seed set $S \subseteq V$ with $|S|\le k$ that maximizes $\sigma_M(S)$ \cite{kempe2003maximizing}. Formally, $\argmax_{S \subseteq V} \sigma_{M}{(S)}, \ \text{such that } \ |S| \leq k.$

\subsection{Pressure Threshold Model Definition}
The PT model is defined on the same graph structure as the LT model and shares the same node activation conditions. Specifically, let \( G = (V, E) \) be a directed graph, where \( V \) denotes the set of nodes and \( E \subseteq V \times V \) denotes the set of directed edges. Each edge $(u, v) \in E$ is assigned a weight $w_{uv} \in [0,1]$ such that for each node $v$, $\sum_{u \in \textcolor{black}{N(v)}} w_{uv} \leq 1$, where $N(v)$ denotes the in-neighbors of $v$. \textcolor{black}{Unless otherwise stated,} each weight $w_{uv}$ \textcolor{black}{is} initialized with a value $1/\text{in-degree}(v)$. Furthermore, each node \( v \in V \) is assigned a threshold \( \theta_v \in (0,1] \), which specifies the minimum cumulative influence required from its incoming neighbors for the node to become active.

The diffusion process unfolds in discrete time steps and begins with an initial seed set \( S \subseteq V \), where the nodes in \( S \) are assumed to be active at time step \( t = 0 \). Let \( A_t \subseteq V \) denote the set of active nodes at time \( t \), with the initial condition \( A_0 = S \). At each subsequent time step, the activation of new nodes is governed by the same rule as the classical Linear Threshold (LT) model: a node \( v \in V \setminus A_t \) becomes active at time \( t+1 \) if the cumulative influence from its currently active in-neighbors meets or exceeds its threshold \( \theta_v \). Formally, the activation condition is given by:
\[
\sum_{u \in P(v)} w_{uv} \geq \theta_v,
\]
where \( P(v) = \{ u \in A_t : (u, v) \in E \} \) denotes the set of active parent nodes (or in-neighbors) of \( v \).

While the activation rule mirrors that of the LT model, the PT model introduces a key modification through a two-phase update mechanism. Specifically, each round of the diffusion process is composed of the following phases:

\begin{enumerate}
    \item \textbf{Activation Phase}: All nodes that satisfy the activation condition based on the current edge weights and active neighbor states become active.
    \item \textbf{Influence Adjustment Phase}: The influence exerted by newly activated nodes on their neighbors is recalibrated based on the 
    influence they 
    received during activation.
\end{enumerate}

\noindent The rationale for the influence adjustment phase is to model a kind of ancestral amplification: a node that was heavily influenced by its neighbors becomes a more influential source in turn. \textcolor{black}{A node's outbound influence is amplified only once during the timestep of its activation, which prevents recursive feedback loops that could otherwise occur in denser regions of the network. Consequently, the adjustment phase updates only edges toward inactive neighbors, since active nodes remain permanently activated, reducing redundant computation and improving efficiency.} To formalize this, let \( v \in V \) be a node that becomes active at time step \( t \). Let \( I_v \) denote the total incoming influence received by \( v \) at the time of its activation:
\[
I_v = \sum_{u \in P(v)} w_{uv}.
\]
The PT model introduces a tunable amplification parameter \( \alpha \geq 0 \), which controls the extent to which this received influence translates into increased outgoing influence. For each newly activated node \( v \) and for each of its outgoing neighbors \( s \) such that \( (v, s) \in E \) and \( \textcolor{black}{s \notin A_t} \), the influence weight is updated according to:
\[
w'_{vs} = \min(1, w_{vs} + \alpha \cdot I_v),
\]
where \( w'_{vs} \) is the new, adjusted influence weight from \( v \) to \( s \). The use of the minimum ensures that the updated weights respect the upper bound of 1, consistent with the range of node thresholds. \textcolor{black}{Because the influence added to each edge is directly proportional to $\alpha I_v$ and bounded by the $\min(1, \cdot)$ constraint, the amplification remains small and localized. The additive form of the amplification scheme reflects the additive nature of the threshold activation function in both the LT and PT models. This design ensures that the model remains consistent with the underlying interpretation of influence aggregation.} Like other standard diffusion models, this process continues until no new nodes can be activated and is progressive, meaning nodes may go from inactive to active, but never the other way. \textcolor{black}{Algorithm~\ref{alg:pressure-threshold} summarizes the complete PT diffusion procedure.}

\textcolor{black}{Conceptually, $\alpha$ serves as a reinforcement coefficient that encodes how strongly social feedback amplifies an individual's influence on their peers. Small values of $\alpha$ correspond to diffusion processes that behave similarly to the Linear Threshold model, while larger values produce stronger social reinforcement dynamics and faster saturation. Because $\alpha$ modulates the intensity of behavioral feedback rather than the network structure itself, in practice it can be treated as a tunable hyperparameter that reflects the degree of social reinforcement present in a given domain and could be estimated or calibrated from empirical diffusion traces, similar to contagion strength in epidemiological models or persuasion intensity in social influence studies \cite{10654694, pmlr-v80-kalimeris18a}.}

\begin{remark}
    As in the LT model, the influence of a seed set \(S \subseteq V\) under the PT model is defined as the expected number of activated nodes by the diffusion process initiated from \(S\), where the expectation is taken over the random thresholds. 
\end{remark}

\begin{algorithm}[tb]
\caption{Pressure Threshold (PT) Diffusion}
\label{alg:pressure-threshold}
\begin{algorithmic}[1]
\Require Directed graph $G=(V,E)$; weights $w_{uv} \in [0,1]$; thresholds $\theta_v \in (0,1]$; seed set $S$; amplification $\alpha \ge 0$
\State \textcolor{black}{$t \gets 0$;\; $A_t \gets S$} \Comment{initial active set}
\Repeat
    \State $N_t \gets \emptyset$ \Comment{newly activated at time $t$}
    \ForAll{$v \in V \setminus A_t$} \Comment{activation phase}
        \State $P_t(v) \gets \{\,u \in A_t : (u,v)\in E\,\}$
        \State $I_v \gets \displaystyle \sum_{u \in P_t(v)} w_{uv}$
        \If{$I_v \ge \theta_v$}
            \State $N_t \gets N_t \cup \{v\}$
        \EndIf
    \EndFor
    \If{\textcolor{black}{$N_t = \emptyset$}}
        \State \textcolor{black}{\textbf{break}} \Comment{\textcolor{black}{no new activations; diffusion halts}}
    \EndIf
    \ForAll{$v \in N_t$} \Comment{influence adjustment phase (one-shot)}
        \ForAll{$(v,s) \in E$ \textbf{with} \textcolor{black}{$s \notin A_t$}} \Comment{only toward currently inactive neighbors}
            \State $w_{vs} \gets \min \bigl(1,\; w_{vs} + \alpha \cdot I_v \bigr)$
        \EndFor
    \EndFor
    \State \textcolor{black}{$A_{t+1} \gets A_t \cup N_t$;\; $t \gets t+1$}
\Until{false}
\State \Return \textcolor{black}{$A_t$}
\end{algorithmic}
\end{algorithm}

\subsection{Computational Complexity}

\begin{theorem}\label{thm:pt-np-hard}
Influence maximization under the Pressure Threshold (PT) diffusion model \textcolor{black}{(PT-IM)} is NP-hard.
\end{theorem}

\begin{proof}
The classical influence maximization problem under the Linear Threshold (LT) model \textcolor{black}{(LT-IM)} is NP-hard \cite{kempe2003maximizing}. We reduce LT-IM to PT-IM in polynomial time. Given an LT instance $(G,k)$, construct the PT instance $(G,k,\alpha=0)$. When $\alpha=0$, the PT process is identical to LT because the adjustment phase is inactive. \textcolor{black}{Formally, for every seed set $S$ we have:
\[
\sigma_{\mathrm{PT}}(S;\alpha=0)=\sigma_{\mathrm{LT}}(S).
\]
Thus, an optimal solution for PT-IM with $\alpha=0$ is an optimal solution for the original LT-IM instance. Thus, any polynomial-time algorithm for PT-IM would yield one for LT-IM, giving a polynomial-time many-one reduction.} Therefore, PT-IM is NP-hard. \qedhere
\end{proof}

\textcolor{black}{\begin{remark}[Decision vs.\ optimization]
The theorem above is stated for the optimization problem (maximizing expected spread). For the decision version (``Does there exist $S\subseteq V$, $|S|\le k$, with expected spread at least $T$?''), the same restriction $\alpha=0$ yields an identical polynomial-time reduction, implying NP-hardness of the decision problem as well.
\end{remark}}

\textcolor{black}{\begin{corollary}[Inapproximability inherited at $\alpha=0$]\label{cor:pt-inapprox}
Under $\alpha=0$, PT reduces to LT. It is NP-hard to approximate optimal influence for LT within a factor better than $1-1/e+\varepsilon$ for any $\varepsilon>0$ unless $\mathrm{P}=\mathrm{NP}$ \cite{kempe2003maximizing}. Consequently, PT-IM inherits the same inapproximability barrier.
\end{corollary}}

This computational complexity result highlights the necessity of efficient approximation or heuristic approaches for practical applications.

\section{Properties of the Proposed Model} \label{sec:model_properties}

\subsection{Model Demonstration}
To build intuition for the reinforcement dynamics introduced by PT, we present a compact visualization contrasting PT and LT on a small controlled instance. The objective is not to claim generality from a toy example, but to provide a clear, step-by-step view of how additive amplification upon activation changes subsequent propagation compared with the classical, static-weight LT evolution on the same graph.

\begin{figure*}[htbp]
\centering
\begin{subfigure}[t]{0.48\textwidth}
    \centering
    \begin{subfigure}[t]{\linewidth}
        \includegraphics[width=\linewidth]{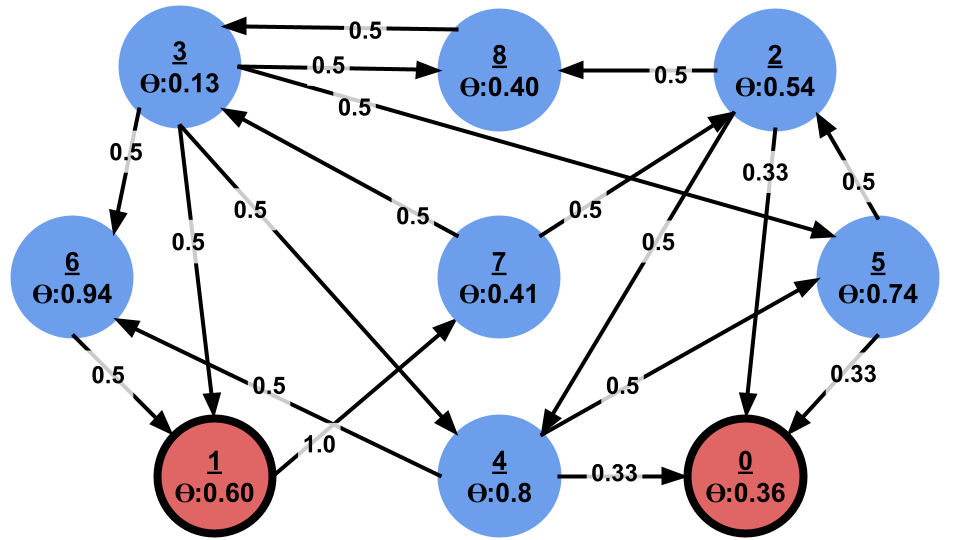}
        \vspace{.5pt}
        \caption*{PT Step 0 (No. of activated nodes = 2)}
        \label{fig:pt-step0}
    \end{subfigure}\vspace{35pt}
    \begin{subfigure}[t]{\linewidth}
        \includegraphics[width=\linewidth]{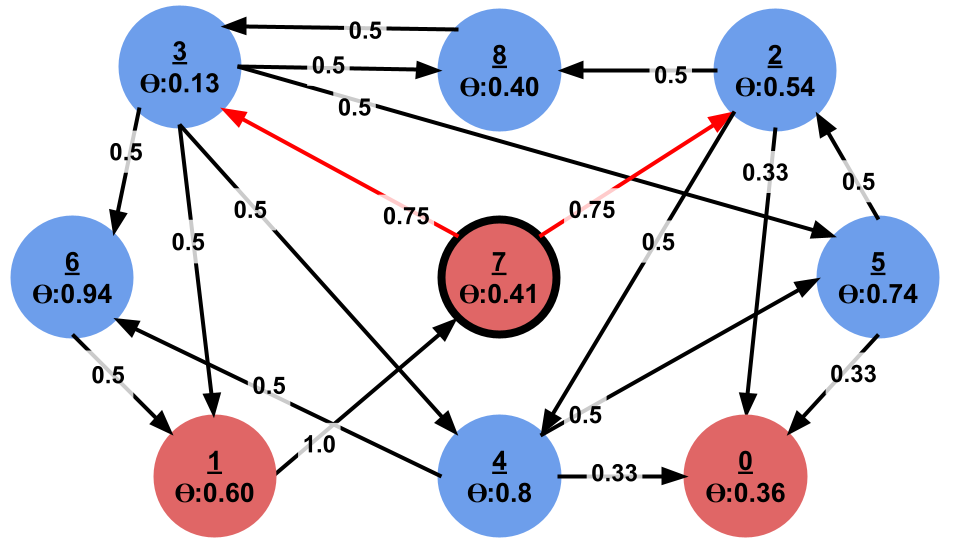}
        \vspace{.5pt}
        \caption*{PT Step 1 (No. of activated nodes = 3)}
        \label{fig:pt-step1}
    \end{subfigure}\vspace{35pt}
    \begin{subfigure}[t]{\linewidth}
        \includegraphics[width=\linewidth]{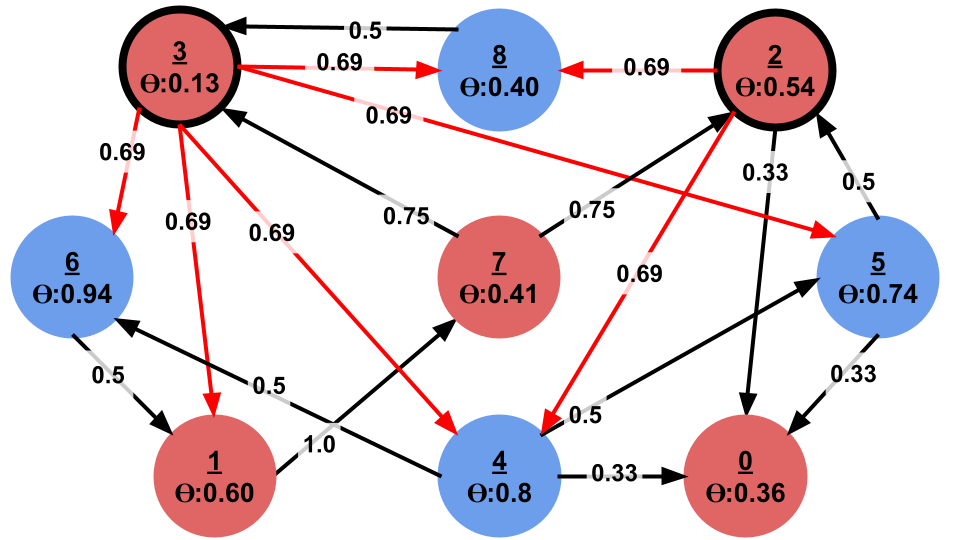}
        \vspace{.5pt}
        \caption*{PT Step 2 (No. of activated nodes = 5)}
        \label{fig:pt-step2}
    \end{subfigure}\vspace{10pt}
\end{subfigure}\hfill
\begin{subfigure}[t]{0.48\textwidth}
    \centering
    \begin{subfigure}[t]{\linewidth}
        \includegraphics[width=\linewidth]{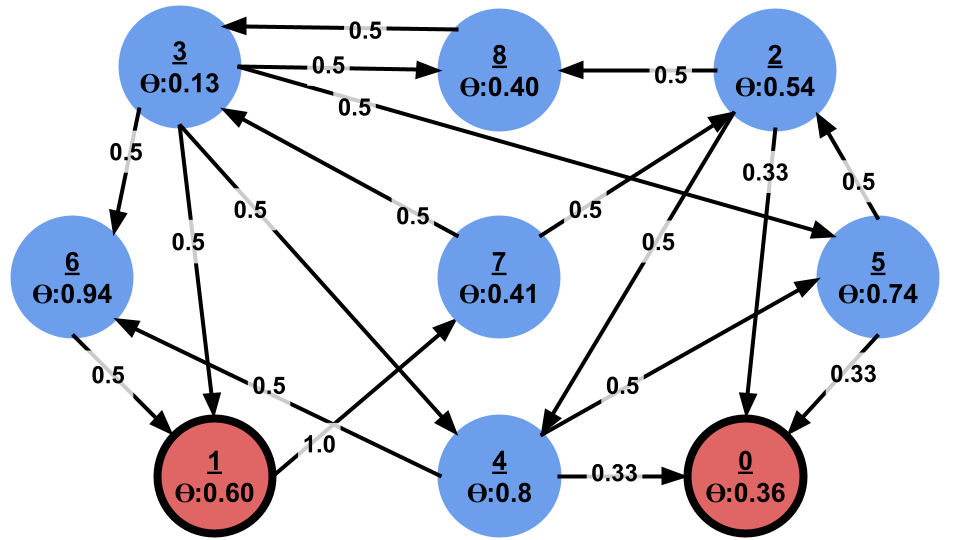}
        \vspace{.5pt}
        \caption*{LT Step 0 (No. of activated nodes = 2)}
        \label{fig:lt-step0}
    \end{subfigure}\vspace{35pt}
    \begin{subfigure}[t]{\linewidth}
        \includegraphics[width=\linewidth]{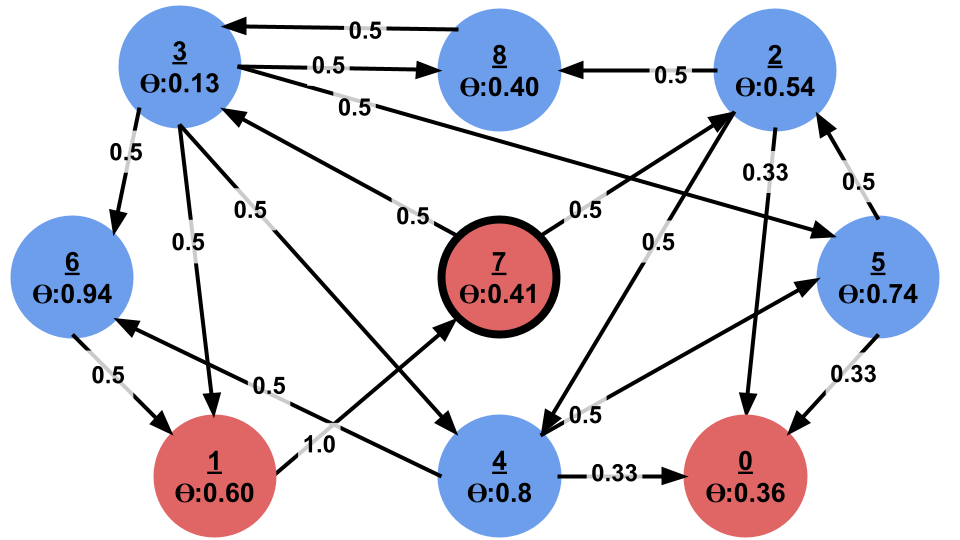}
        \vspace{.5pt}
        \caption*{LT Step 1 (No. of activated nodes = 3)}
        \label{fig:lt-step1}
    \end{subfigure}\vspace{35pt}
    \begin{subfigure}[t]{\linewidth}
        \includegraphics[width=\linewidth]{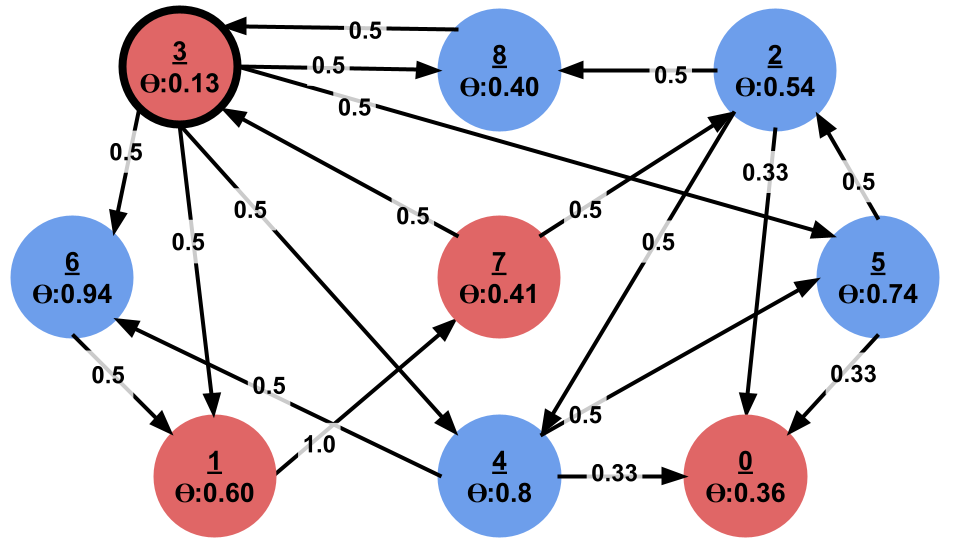}
        \vspace{.5pt}
        \caption*{LT Step 2 (No. of activated nodes = 4)}
        \label{fig:lt-step2}
    \end{subfigure}\vspace{10pt}
\end{subfigure}
\caption{Early diffusion stages under PT and LT on the same graph using an identical seed set. Visualizations show the initial evolution of the diffusion process, during which the reinforcement effects are present but have not yet accumulated strongly. Activated nodes are shown in red, newly activated nodes are highlighted in red with a black outline, and unactivated nodes are shown in blue. Newly amplified edges resulting from PT’s influence adjustment mechanism are highlighted in red. Node labels display the node index (underlined) together with the corresponding activation threshold $\theta$.}
\label{fig:ptlt-early}
\end{figure*}

\begin{figure*}[htbp]
\centering
\begin{subfigure}[t]{0.48\textwidth}
    \centering
    \begin{subfigure}[t]{\linewidth}
        \includegraphics[width=\linewidth]{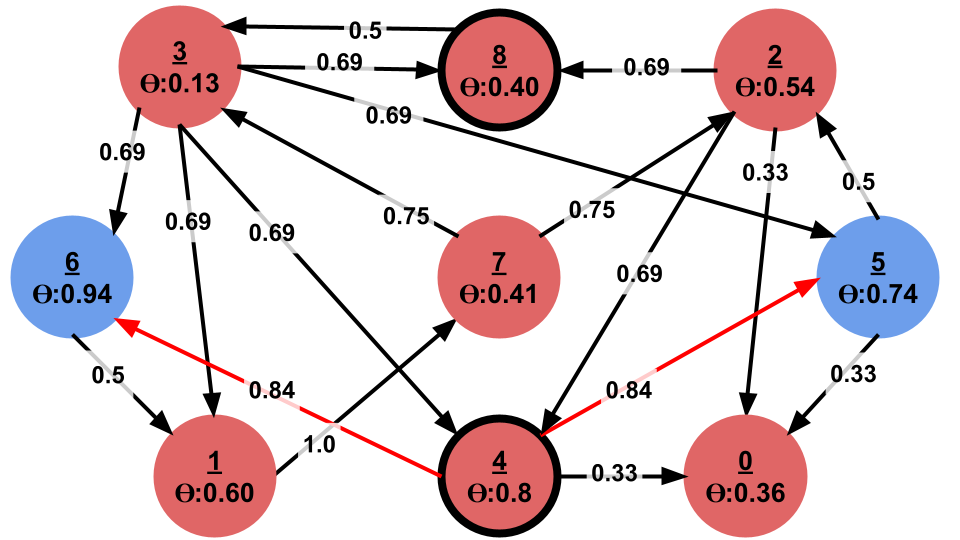}
        \vspace{.5pt}
        \caption*{PT Step 3 (No. of activated nodes = 7)}
        \label{fig:pt-step3}
    \end{subfigure}\vspace{35pt}
    \begin{subfigure}[t]{\linewidth}
        \includegraphics[width=\linewidth]{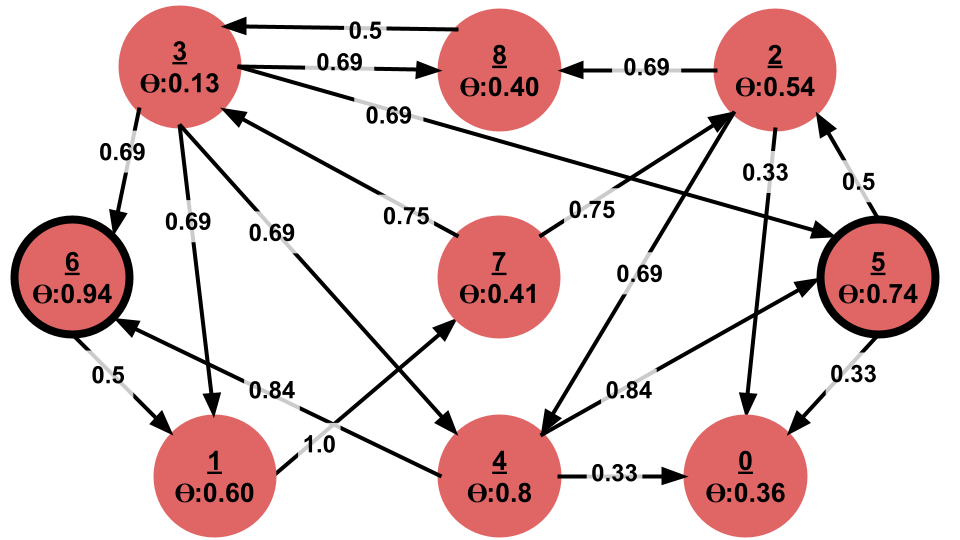}
        \vspace{.5pt}
        \caption*{PT Step 4 (No. of activated nodes = 9)}
        \label{fig:pt-step4}
    \end{subfigure}\vspace{35pt}
    \begin{subfigure}[t]{\linewidth}
        \includegraphics[width=\linewidth]{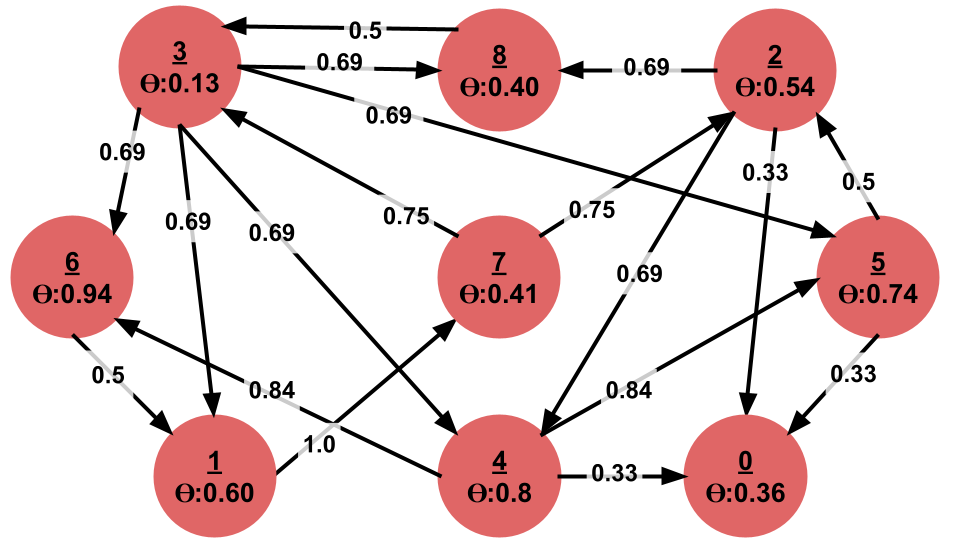}
        \vspace{.5pt}
        \caption*{PT Final Step (No. of activated nodes = 9)}
        \label{fig:pt-step5}
    \end{subfigure}\vspace{10pt}
\end{subfigure}\hfill
\begin{subfigure}[t]{0.48\textwidth}
    \centering
    \begin{subfigure}[t]{\linewidth}
        \includegraphics[width=\linewidth, trim=.05pt .05pt .05pt .05pt, clip]{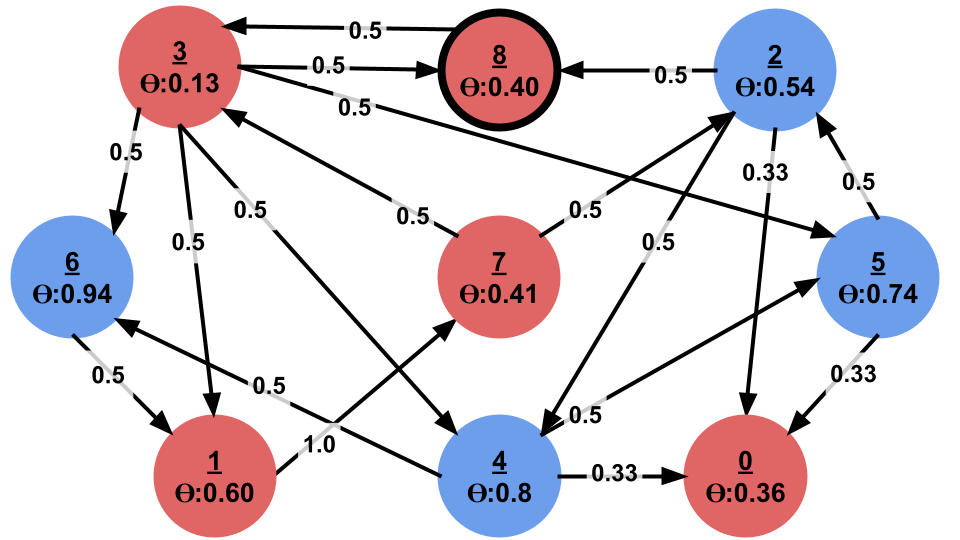}
        \vspace{.5pt}
        \caption*{LT Step 3 (No. of activated nodes = 5)}
        \label{fig:lt-step3}
    \end{subfigure}\vspace{35pt}
    \begin{subfigure}[t]{\linewidth}
        \includegraphics[width=\linewidth, trim=0pt 0pt 0pt 0pt, clip]{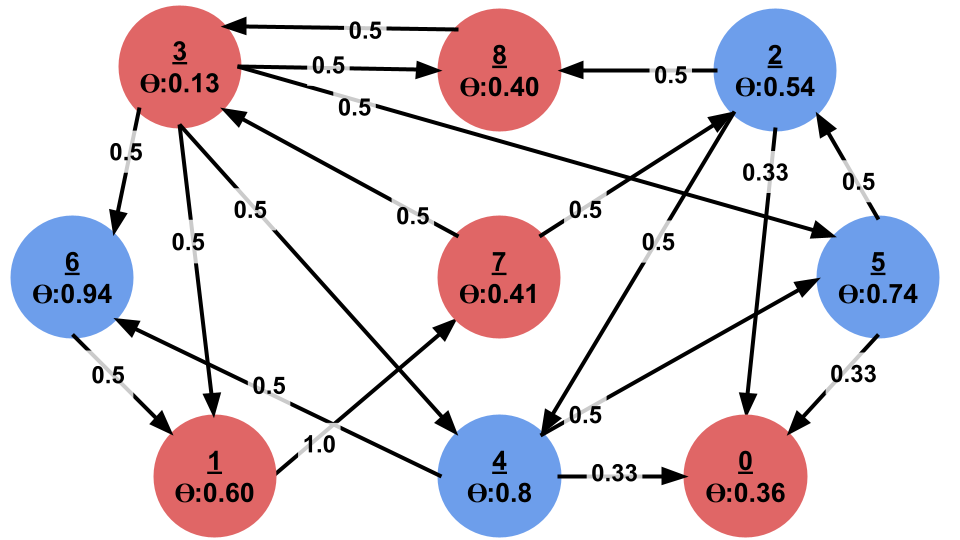}
        \vspace{.5pt}
        \caption*{LT Final Step (No. of activated nodes = 5)}
        \label{fig:lt-final}
    \end{subfigure}\vspace{35pt}
    \begin{subfigure}[t]{\linewidth}
    \end{subfigure}\vspace{10pt}
\end{subfigure}
\caption{Later diffusion stages under PT and LT on the same graph using an identical seed set. As the diffusion evolves, the reinforcement effects intensify, PT continues amplifying outgoing influence along activated edges and attains full coverage, whereas LT halts after activating fewer nodes. Activated nodes are shown in red, newly activated nodes are highlighted in red with a black outline, and unactivated nodes are shown in blue. Newly amplified edges due to PT’s reinforcement mechanism are highlighted in red. Node labels display the node index (underlined) together with the corresponding activation threshold $\theta$.}
\label{fig:ptlt-late}
\end{figure*}

\textcolor{black}{To illustrate the dynamics of the PT model, we present a small-scale example with $\alpha = 0.25$ on a 9-node network. Figures~\ref{fig:ptlt-early} and~\ref{fig:ptlt-late} show the early and later stages of PT diffusion alongside LT on the same graph and seed set. A larger $\alpha$ value was used to better demonstrate the distinction between the two models.}

\textcolor{black}{Both models begin with two identical seeds at $t=0$. Activated nodes are shown in red; nodes newly activated at the current timestep are outlined with a black border; edges whose weights are amplified during the adjustment phase are highlighted in red. As PT evolves, edges incident to newly activated nodes are amplified according to the rule described in Section~\ref{sec:preliminaries}. Using the same seeds, PT 
achieves full coverage, whereas LT terminates with only five activated nodes.}

In the earliest timesteps, PT and LT behave similarly because the initial weights are identical and no adjustment has yet occurred. Differences emerge once the reinforcement mechanism activates: newly activated nodes increase their outgoing influence, allowing neighbors that would remain inactive under LT to cross their thresholds under PT. These incremental adjustments can accumulate in denser regions of the graph and ultimately produce different activation patterns and final reach.

\subsection{Monotonicity Analysis}

While the LT model is known to yield a non-decreasing influence spread function~\cite{kempe2003maximizing}, no analogous guarantee holds under the PT model. 

A set function $f : 2^V \to \mathbb{R}$ is \emph{monotone} if for all $S \subseteq T \subseteq V$,
\[
f(S) \le f(T).
\]

\begin{proposition}\label{prop:pt-not-monotone}
There exists a graph $G$, edge weights, activation thresholds, and a parameter $\alpha > 0$ such that the influence $\sigma_{\mathrm{PT}}(\cdot)$ under the PT model is not monotone.
\end{proposition}

\begin{proof}
Let $G = (V, E)$ with $V = \{a, b, c, d, e\}$ and directed edges with weights as follows: $w_{ca} = w_{da} = 0.5$, $w_{cd} = w_{ed} = w_{bd} = 0.33$, $w_{ab} = w_{db} = 0.5$, and $w_{de} = 1$. Set activation thresholds $\theta_a = 0.1$, $\theta_b = 0.7$, $\theta_d = 0.6$, $\theta_e = 0.8$, and let $\alpha = 0.5$. Define seed sets $S = \{c\}$ and $T = \{a, c\}$, so that $S \subseteq T$.

We compute $\sigma_{\mathrm{PT}}(S)$ and $\sigma_{\mathrm{PT}}(T)$ by tracing the deterministic cascade under each seeding for a fixed threshold realization, as violations need only occur pointwise.

\paragraph{Case $S = \{c\}$.}
Node $c$ is seeded and immediately active. Node $a$ receives influence $w_{ca} = 0.5 > \theta_a$ and activates. Since $a$ was not seeded, PT amplifies the edge $(a, b)$ to $w'_{ab} = 0.75$; as $w'_{ab} > \theta_b$, node $b$ activates. In turn, PT amplifies $(b, d)$ to $w'_{bd} = 0.705$, so $d$ receives cumulative influence $w_{cd} + w'_{bd} = 0.33 + 0.705 = 1.035 > \theta_d$ and activates. Finally, $w'_{de}$ then becomes $min(1, 1 + 0.5 * 1.035) = 1$, and  $e$ receives $w'_{de} = 1 > \theta_e$ and activates. All five nodes are therefore active, giving $\sigma_{\mathrm{PT}}(S) = 5$.

\paragraph{Case $T = \{a, c\}$.}
Both $a$ and $c$ are seeded. Since PT amplification applies only to nodes activated through propagation and not to seeded nodes, $b$ receives only the unmodified weight $w_{ab} = 0.5 < \theta_b$ and remains inactive. Node $d$ receives influence $w_{cd} = 0.33 < \theta_d$ and remains inactive. The final active set is $\{a, c\}$, giving $\sigma_{\mathrm{PT}}(T) = 2$.

Since $S \subseteq T$ yet $\sigma_{\mathrm{PT}}(S) = 5 > 2 = \sigma_{\mathrm{PT}}(T)$, the spread function $\sigma_{\mathrm{PT}}(\cdot)$ violates the monotonicity condition. 
\end{proof}

\begin{remark}
The non-monotonicity in Proposition~\ref{prop:pt-not-monotone} arises from comparing seed sets, not from a single cascade. For any fixed seed set $S$, the active set grows monotonically over time, nodes never deactivate, and $\sigma_{\mathrm{PT}}(S)$ is non-decreasing.
\end{remark}

\subsection{Submodularity Analysis}

While the LT model is known to produce a submodular spread function~\cite{kempe2003maximizing}, the activation-triggered amplification introduced by the PT model can violate the diminishing returns property.

A set function $f : 2^V \to \mathbb{R}$ is \emph{submodular} if for all $S \subseteq T \subseteq V$ and $v \notin T$,
\[
f(S \cup \{v\}) - f(S) \ge f(T \cup \{v\}) - f(T).
\]

\begin{proposition}There exists a graph $G$, edge weights, activation thresholds, and a parameter $\alpha > 0$ such that the influence $\sigma_{\mathrm{PT}}(\cdot)$ under the PT model is not submodular. 

\end{proposition}

\begin{proof}
We use the same graph, weights, thresholds, and $\alpha$ as in Proposition~\ref{prop:pt-not-monotone}, with $S = \{c\}$, $T = \{a,c\}$, and $v = b$. By Proposition~\ref{prop:pt-not-monotone}, $\sigma_{\mathrm{PT}}(S) = 5$ and $\sigma_{\mathrm{PT}}(T) = 2$. For $S \cup \{b\}$: since $b$ is seeded, no amplification is applied to $(b,d)$, so $d$ receives $w_{cd} + w_{bd} = 0.66 > \theta_d$ and activates, as does $e$, giving $\sigma_{\mathrm{PT}}(S \cup \{b\}) = 5$. For $T \cup \{b\}$: similarly, $d$ receives $w_{cd} + w_{bd} = 0.66 > \theta_d$ and all five nodes activate, giving $\sigma_{\mathrm{PT}}(T \cup \{b\}) = 5$. 
The marginal gains are $\sigma_{\mathrm{PT}}(S\cup\{b\})-\sigma_{\mathrm{PT}}(S)=5-5=0$ and $\sigma_{\mathrm{PT}}(T\cup\{b\})-\sigma_{\mathrm{PT}}(T)=5-2=3$.
Since $0 < 3$, the submodularity inequality fails.
\end{proof}

\begin{remark}
Due to pressure effects, the PT influence function $\sigma_{\mathrm{PT}}$ is generally neither monotone nor submodular, and therefore the classical $(1-1/e)$ greedy guarantee \cite{nemhauser1978analysis} does not apply. For non‑monotone submodular maximization under cardinality constraints, the optimal polynomial‑time approximation factor is $1/e$ \cite{feige2011maximizing}. Thus, without monotonicity, one cannot, in general, expect a $(1-1/e)$ guarantee. Nevertheless, it is still informative to quantify how far $\sigma_{\mathrm{PT}}$ departs from submodularity via the submodularity ratio $\gamma_{U,k}(f)$ \cite{approxsubmodular}, defined (for any set function $f$) as
\[
\gamma_{U,k}(f)
\;=\;
\min_{\substack{
L \subseteq U \\
S:\ |S|\le k,\ S\cap L=\varnothing
}}
\frac{\sum_{x\in S}\bigl(f(L\cup\{x\})-f(L)\bigr)}
{f(L\cup S)-f(L)} .
\]
When $\gamma_{U,k}=1$, the function is submodular.

In the PT model, the submodularity ratio satisfies
\[
\gamma_{U,k}(\sigma_{\mathrm{PT}})
\;\ge\;
\underline{\gamma}(G,\alpha),
\qquad
\text{with}
\qquad
\underline{\gamma}(G,\alpha)\xrightarrow[\alpha\to 0]{} 1,
\]
which recovers the classical LT setting and its $(1-1/e)$ approximation guarantee in the zero‑pressure limit. A precise, data-dependent characterization of $\underline{\gamma}(G,\alpha)$ as a function of the graph and pressure is left for future work.
\end{remark}

\section{Experiments}\label{sec:experiments}

\subsection{Network Data}

\textcolor{black}{We evaluate the Pressure Threshold (PT) model against the Linear Threshold (LT) model on four networks spanning distinct topologies and semantics.} Three of these networks are real-world graphs obtained from the Stanford Large Network Dataset Collection (SNAP) \cite{snapnets}: \textcolor{black}{(i) Facebook social circles \cite{leskovec2012learning}, (ii) Wikipedia voting \cite{leskovec2010predicting,leskovec2010signed}, and (iii) Bitcoin OTC trust \cite{kumar2016edge,kumar2018rev2}.} \textcolor{black}{To complement these with a purely synthetic baseline, we also include (iv) an Erd\H{o}s--R\'enyi random network \(G(n,p)\) \cite{erdds1959random}.} The number of nodes, number of directed edges, and edge-to-node ratios are reported in Table~\ref{tab:basic_info}. \textcolor{black}{Throughout, we study progressive cascades (nodes, once activated, remain active) and keep the underlying graph structure fixed during each diffusion run.}

\begin{table}[h] 
    \centering
    \begin{tabular}{l c c c}
        \hline \hline
        Network & Nodes & Edges & Edge/Node Ratio\\
        \hline
        Facebook      & 4{,}039 & 88{,}234  & 21.846 \\
        Bitcoin       & 5{,}881 & 35{,}592  & 6.052  \\
        Wikipedia     & 7{,}115 & 103{,}689 & 14.573 \\
        Erd\H{o}s--R\'enyi  & 5{,}000 & 62{,}597  & 12.519 \\
        \hline
        \hline
    \end{tabular}
    \caption{Basic statistics of the networks used in experiments.}
    \label{tab:basic_info}
\end{table}

\paragraph{Graph construction and directionality.}
The Facebook circles dataset is originally undirected, with an edge indicating a mutual friendship relation. \textcolor{black}{Following standard practice in LT-based diffusion studies \cite{kempe2003maximizing}, we convert each undirected edge $\{u,v\}$ into two directed edges $(u,v)$ and $(v,u)$ to preserve symmetric influence opportunities.} The Bitcoin OTC trust network is inherently directed: an edge $(u,v)$ represents a rating or trust assignment from user $u$ to user $v$ on a peer-to-peer trading platform. The Wikipedia voting network is also directed: \textcolor{black}{an edge $(u,v)$ indicates that user $u$ voted for user $v$ in an admin election, producing a directed endorsement flow aligned with influence propagation.} \textcolor{black}{Similar to the Facebook network, for the Erd\H{o}s--R\'enyi random graph, we generate $G(n,p)$ with $n=5{,}000$ and $p=0.005$, and convert each undirected edge into two directed arcs.}

\paragraph{Edge weights (weighted-cascade normalization).}
Unless stated otherwise, we use the weighted-cascade convention \cite{kempe2003maximizing}, assigning for each directed edge \((u,v)\) the weight:
\[
w_{uv} \;=\; \frac{1}{\text{in-degree}(v)}.
\]
This ensures that for every node \(v\), the incoming weights sum to one, i.e., \(\sum_{u\in N(v)} w_{uv} = 1\). \textcolor{black}{In the PT model, these weights may be updated according to the additive adjustment rule upon activation events (Section~\ref{sec:preliminaries}), while respecting the upper bound \(w'_{uv} \le 1\) via the \(\min(1,\cdot)\) cap.}

\paragraph{Thresholds and randomness.}
In all runs, node thresholds \(\theta_v\) are drawn i.i.d.\ uniformly from \((0,1]\) at the beginning of each Monte Carlo (MC) simulation, consistent with the LT literature. \textcolor{black}{For paired comparisons (PT vs.\ LT) under the same seed budget \(k\), we reuse the same threshold instantiation to isolate model effects from sampling noise.} \textcolor{black}{All reported spreads are averages over independent MC trials, with the number of trials specified per experiment.}

\subsection{Experiment Details}\label{subsec:exp_details}

\textcolor{black}{We perform three complementary experiments to probe (i) the qualitative effect of PT’s reinforcement on seed selection (Experiment~1), (ii) the quantitative impact on influence spread versus budget across heterogeneous graphs (Experiment~2), and (iii) the sensitivity of PT to the reinforcement parameter $\alpha$ (Experiment~3). We use the CELF algorithm \cite{leskovec2007cost} for seed selection in Experiments~1 and~2 to maintain consistency with standard IM practice.}

\subsubsection{Experiment 1: Seed selection under PT vs.\ LT.}
The goal of this experiment is to examine whether PT and LT identify the same influential nodes when given the same budget \(k\). We run CELF for the Facebook network with budget \(k=20\). To evaluate the marginal gains within CELF, we estimate spreads via 1{,}000 MC simulations per evaluation. \textcolor{black}{For the PT runs, we set $\alpha=0.1$, which empirically balances (i) observable reinforcement leading to altered seed rankings and (ii) avoidance of near-instant saturation.} For fairness, the same set of random threshold instantiations is shared between LT and PT whenever possible, ensuring that any observed differences are attributable to the model and not to randomness. The order of seed node selection is preserved for side-by-side comparison.

\subsubsection{Experiment 2: Influence vs.\ budget.}
To quantify the downstream effect of PT’s reinforcement, we compute the influence spread \(\sigma(S)\) as a function of the seed budget \(k\) across the four graphs in Table~\ref{tab:basic_info}. For each network and each budget \(k\in\{1,2,\dots,60\}\), we run CELF under three diffusion settings: LT, PT with \(\alpha=0.001\), and PT with \(\alpha=0.005\). Each evaluation of \(\sigma(S)\) within CELF averages 1{,}000 MC simulations. \textcolor{black}{These $\alpha$ values were selected empirically to illustrate qualitative differences between PT and LT while ensuring stable and interpretable diffusion dynamics. Pilot tests with larger $\alpha$ rapidly saturated the denser graphs (Facebook, Wikipedia), while substantially smaller $\alpha$ yielded behavior nearly identical to LT.} 

\begin{remark}
    We also tested Normal$(0.5,0.1)$, Beta$(2,5)$, and Beta$(5,2)$ threshold priors; all produced PT–LT spread gaps similar to the uniform prior, so redundant plots are omitted.
\end{remark}

\subsubsection{Experiment 3: Sensitivity of PT to \(\alpha\).}
We probe how PT’s final spread scales with the reinforcement parameter by fixing a seed set size (\(k=10\)) on Facebook and sweeping \(\alpha\) over a fine grid \([10^{-4}, 10^{-1}]\) with step \(10^{-4}\). For each \(\alpha\), we perform 1{,}000 MC simulations and report the average spread. \textcolor{black}{To reduce high-frequency variability due to stochastic thresholds, we apply a centered moving average with a window of 9 grid points; each plotted value is the average of the focal $\alpha$ and its four nearest neighbors.}

\subsection{Results}

\begin{figure*}[htbp]
\centering
\includegraphics[width=2.\columnwidth,height=4.9cm]{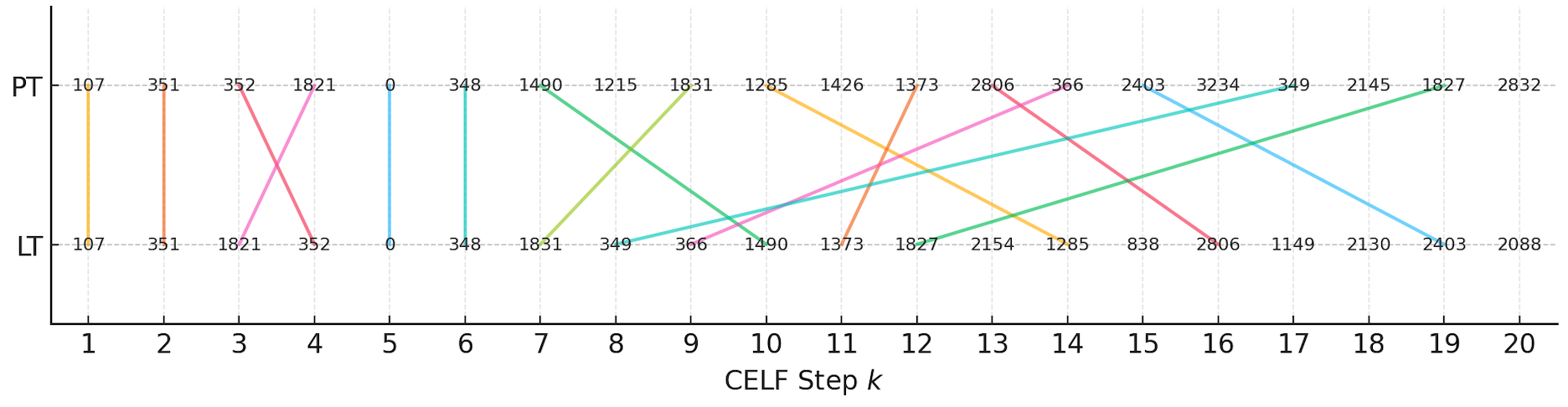}
\caption{CELF seed-selection timeline for the Facebook network (PT vs.\ LT, \(k=20\)). Horizontal position shows the selection step; vertical links mark common seeds. Strong early overlap is present, with divergence emerging as PT favors reinforcement.}
\label{fig:timeline}
\end{figure*}

The results for Experiment~1 appear in Table~\ref{tab:timeline} and Figure~\ref{fig:timeline}. The results for Experiment~2 are shown in Figures~\ref{fig:FB}--\ref{fig:Rand}. The results for Experiment~3 are given in Figure~\ref{fig:alpha}.

\begin{table}[h]
\centering
\begin{tabular}{p{0.40\columnwidth} p{0.49\columnwidth}}
\hline\hline
Model & Seeds selected (in order) \\
\hline
Linear Threshold   & 107, 351, 1821, 352, 0, 348, 1831, 349, 366, 1490, 1373, 1827, 2154, 1285, 838, 2806, 1149, 2130, 2403, 2088 \\
\hline
Pressure Threshold & 107, 351, 352, 1821, 0, 348, 1490, 1215, 1831, 1285, 1426, 1373, 2806, 366, 2403, 3234, 349, 2145, 1827, 2832 \\
\hline
\hline
\end{tabular}
\caption{Ordered seed sets returned by CELF for the Facebook network \(k=20\). The overlap in early positions and divergence in later positions reveal how PT’s reinforcement can elevate otherwise secondary vertices.}
\label{tab:timeline}
\end{table}

\begin{figure}[h]
\centering
\resizebox{.87\columnwidth}{!}{%
    \begin{tikzpicture}
\definecolor{color0}{rgb}{0.12156862745098,0.466666666666667,0.705882352941177}
\definecolor{color1}{rgb}{1,0.498039215686275,0.0549019607843137}
\definecolor{color2}{rgb}{0.172549019607843,0.627450980392157,0.172549019607843}
\begin{axis}[
legend cell align={left},
reverse legend,
legend style={
  fill opacity=0.8,
  draw opacity=1,
  text opacity=1,
  at={(0.97,0.25)},
  anchor=east,
  draw=white!80!black
},
tick align=outside,
tick pos=left,
x grid style={white!69.0196078431373!black},
xlabel={Budget \(\displaystyle k\)},
xmajorgrids,
xmin=0, xmax=61,
xtick style={color=black},
y grid style={white!69.0196078431373!black},
ylabel={Influence},
ymajorgrids,
ymin=0, ymax=3600,
ytick style={color=black}
]
\addplot [thick, color0, mark=*, mark size=.5, mark options={solid}]
table {%
1 334.642
2 608.443
3 833.666
4 1020.248
5 1181.6
6 1262.681
7 1334.467
8 1384.669
9 1430.347
10 1472.063
11 1515.548
12 1551.514
13 1589.55
14 1626.002
15 1656.765
16 1687.273
17 1713.757
18 1739.178
19 1763.244
20 1787.961
21 1812.06
22 1836.995
23 1858.174
24 1879.514
25 1901.686
26 1922.769
27 1942.921
28 1968.146
29 1989.563
30 2009.174
31 2028.747
32 2047.13
33 2066.004
34 2084
35 2101.559
36 2118.451
37 2135.81
38 2152.409
39 2169.873
40 2186.079
41 2202.106
42 2217.497
43 2232.437
44 2249.494
45 2264.231
46 2278.804
47 2292.923
48 2307.515
49 2322.161
50 2335.654
51 2348.992
52 2361.98
53 2374.705
54 2387.533
55 2400.321
56 2412.639
57 2425.011
58 2437.552
59 2449.382
60 2461.219
};
\addlegendentry{Linear Threshold ($\alpha=0$)}
\addplot [thick, color1, mark=*, mark size=.5, mark options={solid}]
table {%
1 391.72
2 688.303
3 922.964
4 1151.397
5 1312.088
6 1400.988
7 1472.899
8 1537.7
9 1592.164
10 1645.537
11 1695.249
12 1743.071
13 1790.783
14 1834.418
15 1873.344
16 1909.028
17 1946.952
18 1981.673
19 2015.471
20 2044.356
21 2073.682
22 2101.406
23 2131.614
24 2158.072
25 2184.315
26 2208.925
27 2233.526
28 2259.445
29 2284.054
30 2307.706
31 2330.674
32 2352.734
33 2373.766
34 2394.842
35 2414.687
36 2433.409
37 2452.353
38 2471.973
39 2491.125
40 2508.848
41 2525.745
42 2545.317
43 2562.675
44 2579.178
45 2597.966
46 2613.8
47 2631.358
48 2645.633
49 2661.032
50 2675.743
51 2689.133
52 2703.465
53 2717.398
54 2731.408
55 2744.001
56 2756.562
57 2769.202
58 2782.859
59 2797.137
60 2810.784
};
\addlegendentry{Pressure Threshold ($\alpha=0.001$)}
\addplot [thick, color2, mark=*, mark size=.5, mark options={solid}]
table {%
1 660.579
2 1078.835
3 1455.629
4 1727.461
5 1907.756
6 2033.591
7 2137.707
8 2222.465
9 2310.523
10 2376.827
11 2439.832
12 2497.333
13 2545.548
14 2585.35
15 2629.415
16 2666.051
17 2703.345
18 2739.729
19 2775.337
20 2810.059
21 2839.22
22 2867.473
23 2896.07
24 2922.519
25 2947.255
26 2974.572
27 2999.281
28 3023.28
29 3046.049
30 3067.716
31 3090.573
32 3110.024
33 3130.296
34 3150.115
35 3169.333
36 3190.846
37 3210.373
38 3225.988
39 3243.655
40 3258.229
41 3272.752
42 3287.554
43 3301.066
44 3314.712
45 3329.657
46 3344.561
47 3356.026
48 3368.143
49 3379.119
50 3393.093
51 3403.333
52 3413.243
53 3423.035
54 3432.982
55 3445.419
56 3455.29
57 3464.391
58 3473.042
59 3481.419
60 3490.28
};
\addlegendentry{Pressure Threshold ($\alpha = 0.005$)}
\end{axis}
\end{tikzpicture}
}
\caption{Influence vs.\ budget for the Facebook network. PT outperforms LT, especially for larger \(\alpha\).}
\label{fig:FB}
\end{figure}

\begin{figure}[h]
\centering
\resizebox{.87\columnwidth}{!}{%
    \begin{tikzpicture}
\definecolor{color0}{rgb}{0.12156862745098,0.466666666666667,0.705882352941177}
\definecolor{color1}{rgb}{1,0.498039215686275,0.0549019607843137}
\definecolor{color2}{rgb}{0.172549019607843,0.627450980392157,0.172549019607843}
\begin{axis}[
legend cell align={left},
reverse legend,
legend style={
  fill opacity=0.8,
  draw opacity=1,
  text opacity=1,
  at={(0.97,0.25)},
  anchor=east,
  draw=white!80!black
},
tick align=outside,
tick pos=left,
x grid style={white!69.0196078431373!black},
xlabel={Budget \(\displaystyle k\)},
xmajorgrids,
xmin=0, xmax=61,
xtick style={color=black},
y grid style={white!69.0196078431373!black},
ylabel={Influence},
ymajorgrids,
ymin=0, ymax=7500,
ytick style={color=black}
]
\addplot [thick, color0, mark=*, mark size=.5, mark options={solid}]
table {%
1 724.446
2 1169.028
3 1565.53
4 1921.167
5 2240.728
6 2490.144
7 2710.867
8 2883.476
9 3057.97
10 3220.825
11 3354.994
12 3481.557
13 3592.957
14 3695.593
15 3801.452
16 3891.044
17 3978.995
18 4068.885
19 4151.053
20 4228.357
21 4300.316
22 4374.764
23 4440.992
24 4502.643
25 4560.075
26 4613.92
27 4669.627
28 4730.352
29 4780.217
30 4829.166
31 4874.286
32 4916.201
33 4959.251
34 5000.799
35 5043.995
36 5080.665
37 5116.117
38 5152.02
39 5189.203
40 5221.273
41 5254.385
42 5286.392
43 5315.155
44 5344.004
45 5372.311
46 5401.982
47 5429.512
48 5458.084
49 5485.69
50 5511.501
51 5536.72
52 5560.478
53 5582.978
54 5606.244
55 5627.405
56 5648.86
57 5668.739
58 5689.415
59 5707.882
60 5725.608
};
\addlegendentry{Linear Threshold ($\alpha=0$)}
\addplot [thick, color1, mark=*, mark size=.5, mark options={solid}]
table {%
1 1054.106
2 1922.544
3 2697.091
4 3393.084
5 3958.991
6 4410.341
7 4726.157
8 5012.502
9 5279.115
10 5500.902
11 5661.572
12 5825.633
13 5987.494
14 6124.452
15 6269.464
16 6372.591
17 6466.473
18 6549.754
19 6620.768
20 6688.814
21 6760.987
22 6827.113
23 6879.977
24 6920.436
25 6957.018
26 6986.451
27 7017.48
28 7051.771
29 7082.021
30 7105.058
31 7115
32 7115
33 7115
34 7115
35 7115
36 7115
37 7115
38 7115
39 7115
40 7115
41 7115
42 7115
43 7115
44 7115
45 7115
46 7115
47 7115
48 7115
49 7115
50 7115
51 7115
52 7115
53 7115
54 7115
55 7115
56 7115
57 7115
58 7115
59 7115
60 7115
};
\addlegendentry{Pressure Threshold ($\alpha=0.001$)}
\addplot [thick, color2, mark=*, mark size=.5, mark options={solid}]
table {%
1 4773.406
2 5949.983
3 6604.243
4 6961.594
5 7115
6 7115
7 7115
8 7115
9 7115
10 7115
11 7115
12 7115
13 7115
14 7115
15 7115
16 7115
17 7115
18 7115
19 7115
20 7115
21 7115
22 7115
23 7115
24 7115
25 7115
26 7115
27 7115
28 7115
29 7115
30 7115
31 7115
32 7115
33 7115
34 7115
35 7115
36 7115
37 7115
38 7115
39 7115
40 7115
41 7115
42 7115
43 7115
44 7115
45 7115
46 7115
47 7115
48 7115
49 7115
50 7115
51 7115
52 7115
53 7115
54 7115
55 7115
56 7115
57 7115
58 7115
59 7115
60 7115
};
\addlegendentry{Pressure Threshold ($\alpha=0.005$)}
\end{axis}
\end{tikzpicture}
}
\caption{Influence vs.\ budget for the Wikipedia network, with PT consistently exceeding LT.}
\label{fig:Wiki}
\end{figure}

\begin{figure}[h]
\centering
\resizebox{.87\columnwidth}{!}{%
    \begin{tikzpicture}
\definecolor{color0}{rgb}{0.12156862745098,0.466666666666667,0.705882352941177}
\definecolor{color1}{rgb}{1,0.498039215686275,0.0549019607843137}
\definecolor{color2}{rgb}{0.172549019607843,0.627450980392157,0.172549019607843}
\begin{axis}[
legend cell align={left},
reverse legend,
legend style={
  fill opacity=0.8,
  draw opacity=1,
  text opacity=1,
  at={(0.97,0.25)},
  anchor=east,
  draw=white!80!black
},
tick align=outside,
tick pos=left,
x grid style={white!69.0196078431373!black},
xlabel={Budget \(\displaystyle k\)},
xmajorgrids,
xmin=0, xmax=61,
xtick style={color=black},
y grid style={white!69.0196078431373!black},
ylabel={Influence},
ymajorgrids,
ymin=0, ymax=5200,
ytick style={color=black}
]
\addplot [thick, color0, mark=*, mark size=.5, mark options={solid}]
table {%
1 828.423
2 1210.393
3 1525.546
4 1798.743
5 2012.038
6 2192.477
7 2359.435
8 2510.854
9 2655.279
10 2766.19
11 2866.307
12 2963.699
13 3061.837
14 3144.634
15 3228.139
16 3303.154
17 3368.111
18 3427.513
19 3485.074
20 3538.395
21 3591.475
22 3640.215
23 3684.06
24 3728.124
25 3771.831
26 3816.506
27 3861.598
28 3898.594
29 3933.908
30 3967.226
31 3999.591
32 4032.339
33 4063.467
34 4092.953
35 4120.236
36 4147.395
37 4173.717
38 4197.457
39 4218.886
40 4240.268
41 4261.696
42 4283.131
43 4303.098
44 4323.809
45 4341.36
46 4360.538
47 4378.264
48 4394.572
49 4409.869
50 4425.074
51 4440.675
52 4455.477
53 4470.959
54 4484.948
55 4498.833
56 4512.495
57 4526.011
58 4539.067
59 4552.136
60 4565.969
};
\addlegendentry{Linear Threshold ($\alpha=0$)}
\addplot [thick, color1, mark=*, mark size=.5, mark options={solid}]
table {%
1 864.599
2 1274.755
3 1610.443
4 1890.631
5 2117.072
6 2307.742
7 2478.306
8 2631.372
9 2781.393
10 2895.744
11 3002.54
12 3102.612
13 3198.584
14 3284.893
15 3367.647
16 3440.966
17 3505.275
18 3565.901
19 3623.79
20 3674.454
21 3725.641
22 3772.414
23 3816.699
24 3860.671
25 3904.088
26 3950.613
27 3992.299
28 4027.321
29 4061.578
30 4094.259
31 4127.623
32 4158.537
33 4191.304
34 4217.246
35 4243.633
36 4268.515
37 4293.71
38 4317.123
39 4339.077
40 4360.589
41 4380.307
42 4399.322
43 4417.985
44 4434.989
45 4452.783
46 4469.303
47 4485.665
48 4503.364
49 4519.749
50 4534.898
51 4549.35
52 4564.905
53 4580.528
54 4594.44
55 4608.196
56 4622.271
57 4635.627
58 4649.094
59 4662.315
60 4673.804
};
\addlegendentry{Pressure Threshold ($\alpha=0.001$)}
\addplot [thick, color2, mark=*, mark size=.5, mark options={solid}]
table {%
1 1083.096
2 1592.946
3 2023.862
4 2344.98
5 2598.107
6 2823.061
7 3005.649
8 3191.511
9 3334.292
10 3463.046
11 3574.688
12 3677.572
13 3772.426
14 3850.951
15 3918.337
16 3981.264
17 4041.128
18 4091.113
19 4136.607
20 4180.012
21 4225.34
22 4266.772
23 4308.036
24 4348.923
25 4384.844
26 4418.224
27 4450.732
28 4481.898
29 4511.866
30 4539.837
31 4563.312
32 4589.278
33 4610.818
34 4631.291
35 4649.316
36 4666.999
37 4684.361
38 4700.332
39 4716.425
40 4732.569
41 4748.532
42 4763.632
43 4780.511
44 4794.649
45 4808.452
46 4821.548
47 4835.095
48 4849.015
49 4862.436
50 4874.262
51 4885.248
52 4896.737
53 4908.094
54 4918.627
55 4928.907
56 4938.522
57 4948.387
58 4958.293
59 4968.401
60 4977.789
};
\addlegendentry{Pressure Threshold ($\alpha=0.005$)}
\end{axis}
\end{tikzpicture}
}
\caption{Influence vs.\ budget for the Bitcoin network. PT gains are modest due to network sparsity.}
\label{fig:BC}
\end{figure}

\begin{figure}[h]
\centering
\resizebox{.87\columnwidth}{!}{%
    \begin{tikzpicture}
\definecolor{color0}{rgb}{0.12156862745098,0.466666666666667,0.705882352941177}
\definecolor{color1}{rgb}{1,0.498039215686275,0.0549019607843137}
\definecolor{color2}{rgb}{0.172549019607843,0.627450980392157,0.172549019607843}
\begin{axis}[
legend cell align={left},
reverse legend,
legend style={
  fill opacity=0.8,
  draw opacity=1,
  text opacity=1,
  at={(0.97,0.25)},
  anchor=east,
  draw=white!80!black
},
tick align=outside,
tick pos=left,
x grid style={white!69.0196078431373!black},
xlabel={Budget \(\displaystyle k\)},
xmajorgrids,
xmin=0, xmax=61,
xtick style={color=black},
y grid style={white!69.0196078431373!black},
ylabel={Influence},
ymajorgrids,
ymin=0, ymax=5500,
ytick style={color=black}
]
\addplot [thick, color0, mark=*, mark size=.5, mark options={solid}]
table {%
1 236.911
2 496.616
3 702.837
4 876.178
5 1067.877
6 1232.833
7 1392.918
8 1546.086
9 1703.67
10 1847.804
11 1973.792
12 2105.853
13 2240.152
14 2368.532
15 2484.657
16 2599.259
17 2727.107
18 2824.584
19 2921.274
20 3011.234
21 3102.095
22 3214.637
23 3294.092
24 3387.032
25 3461.638
26 3546.502
27 3615.883
28 3694.263
29 3757.151
30 3834.451
31 3897.627
32 3957.677
33 4018.885
34 4084.961
35 4141.081
36 4197.002
37 4254.069
38 4317.531
39 4367.534
40 4418.623
41 4468.527
42 4516.362
43 4561.494
44 4612.05
45 4656.926
46 4701.074
47 4740.73
48 4779.572
49 4816.236
50 4852.233
51 4888.328
52 4921.855
53 4954.63
54 4988.449
55 5000.395
56 5000.381
57 5000.314
58 5000.528
59 5000.517
60 5000.673
};
\addlegendentry{Linear Threshold ($\alpha=0$)}
\addplot [thick, color1, mark=*, mark size=.5, mark options={solid}]
table {%
1 301.215
2 554.528
3 773.746
4 1007.202
5 1218.937
6 1403.755
7 1593.759
8 1789.017
9 1956.993
10 2129.5
11 2281.41
12 2438.065
13 2599.626
14 2744.491
15 2878.351
16 3012.517
17 3131.448
18 3253.815
19 3378.235
20 3508.045
21 3614.405
22 3714.713
23 3817.254
24 3919.004
25 4006.886
26 4090.31
27 4179.237
28 4259.916
29 4333.491
30 4406.899
31 4477.306
32 4549.979
33 4617.915
34 4680.458
35 4740.727
36 4800.141
37 4856.382
38 4910.448
39 4961.671
40 5000.792
41 5000.805
42 5000.75
43 5000.11
44 5000.744
45 5000.891
46 5000.681
47 5000.209
48 5000.999
49 5000.928
50 5000.607
51 5000.577
52 5000.774
53 5000.228
54 5000.876
55 5000.846
56 5000.896
57 5000.017
58 5000.106
59 5000.862
60 5000.469
};
\addlegendentry{Pressure Threshold ($\alpha=0.001$)}
\addplot [thick, color2, mark=*, mark size=.5, mark options={solid}]
table {%
1 409.414
2 767.34
3 1103.733
4 1442.252
5 1751.58
6 2033.052
7 2307.674
8 2537.749
9 2762.884
10 2985.4
11 3213.917
12 3401.161
13 3593.41
14 3775.471
15 3960.66
16 4127.566
17 4289.577
18 4432.05
19 4555.37
20 4678.688
21 4783.509
22 4887.983
23 4985.327
24 5000
25 5000
26 5000
27 5000
28 5000
29 5000
30 5000
31 5000
32 5000
33 5000
34 5000
35 5000
36 5000
37 5000
38 5000
39 5000
40 5000
41 5000
42 5000
43 5000
44 5000
45 5000
46 5000
47 5000
48 5000
49 5000
50 5000
51 5000
52 5000
53 5000
54 5000
55 5000
56 5000
57 5000
58 5000
59 5000
60 5000
};
\addlegendentry{Pressure Threshold ($\alpha=0.005$)}
\end{axis}
\end{tikzpicture}
}
\caption{Influence vs.\ budget for the Erd\H{o}s--R\'enyi random network. Amplification accelerates full coverage.}
\label{fig:Rand}
\end{figure}

\begin{figure}[h]
\centering
\resizebox{.87\columnwidth}{!}{%
    \input{figures/Varying_Alpha}
}
\caption{Effect of \(\alpha\) on PT influence (smoothed using a centered moving average with window size of 9) for the Facebook network (\(k=10\)); influence increases monotonically 
with reinforcement strength.
}
\label{fig:alpha}
\end{figure}

\subsection{Discussion}

Figure~\ref{fig:timeline} compares the seed sets that CELF selected under the LT and PT models for a budget of \(k=20\) for the Facebook network. The first column lists the LT seeds, whereas the second lists the PT seeds. Lines connect the vertices common to both seed sets. This graph shows that solving the IM problem under PT yields different seed sets than LT, highlighting a key distinction between the models. The models agree on most early selections (those with the largest marginal gains), but diverge more and more in ordering and in selection as the iteration proceeds. From step~8 onward, PT introduces vertices (e.g., 1215) that never appear in the LT list, indicating that amplification can elevate otherwise secondary nodes. Conversely, several LT‑specific vertices never surface under PT. Because PT applies an additive effect to edge weights rather than a diminishing one, these omissions imply that other vertices experience larger marginal‑gain increases and overtake them in priority. \textcolor{black}{We also observe that the PT model’s seed selection tends to favor nodes embedded within denser regions of the network, where local clustering and mutual reinforcement create more opportunities for amplification compared to nodes with large but sparsely connected neighborhoods. This behavior arises because the PT process implicitly places greater value on a node’s in-degree, reflecting the cumulative pressure it can receive from active neighbors rather than solely on out-degree, as in traditional LT and IC models.} Table \ref{tab:timeline} presents the ordered lists in compact form.

Figures~\ref{fig:FB}--\ref{fig:Rand} show the influence of diffusions across the four networks as the budget varies. As expected, network diffusions with the PT model result in a larger influence spread than the LT model. What is interesting about this experiment is the margin of variation under the PT model between the different network types. For example, the Facebook and Wikipedia networks result in much larger influence spreads for the PT model than the LT model when compared to the Bitcoin network, whose influence under the PT model is not much larger than under the LT model. This is most likely due to the larger edge-to-node ratios of the former, as given in Table \ref{tab:basic_info}. Intuitively, this makes sense, as more edges yield more opportunity for the influence amplification to take place (phase 2 in the PT diffusion). The amplification was so large that some networks were fully diffused before reaching the max budget of \(k = 60\), such as the Erd\H{o}s--R\'enyi network reaching full coverage at \(k = 22\). Despite these scale differences, the influence curves for LT and PT maintain similar shapes, suggesting that PT adds an approximately constant offset to the baseline diffusion rate. The logarithmic shape of the PT curves, which matches the diminishing returns of the LT curve, also suggests practical submodularity of the influence spread function on the datasets tested.

Figure \ref{fig:alpha} shows the influence spread of a PT diffusion for the Facebook network as a function of the parameter \(\alpha\). Because of the large amount of data points, we applied a centered moving average with a window size of 9 to smooth the \(\alpha\)-vs-influence curve. In the plot, each point is replaced by the average of the point itself and the four points to its right and left. As expected, increasing the \(\alpha\) value increases the influence amplification at each node, which increases the total influence spread. The fit can be approximated with the following cubic polynomial function (for descriptive purposes only, not as a predictive or mechanistic model):
\begin{align*}
\text{Influence}(\alpha) &\approx
  1.358 \times 10^{6}\,\alpha^{3} - 6.255 \times 10^{5}\,\alpha^{2} \\
&\quad + 8.485 \times 10^{4}\,\alpha
  + 57.29.
\end{align*}

\section{Conclusion} \label{sec:conclusion}

\subsection{Summary of \textcolor{black}{Contributions}}
In this work, we introduced the Pressure Threshold (PT) model, a diffusion framework that captures reinforcement effects and cumulative influence in social networks. PT extends the Linear Threshold (LT) model by amplifying a node’s outgoing influence according to the pressure it receives from neighbors. We showed that influence maximization under PT remains NP-hard and that PT is neither monotone nor submodular when $\alpha>0$, implying that classical greedy guarantees may not fully apply. Experiments on real and synthetic networks demonstrated that PT yields seed sets and spreads distinct from LT, with stronger gains in denser networks and more modest effects in sparse ones. These findings highlight PT’s ability to model reinforcement-driven diffusion not captured by LT. 

\subsection{Real-World Implications}
The Pressure Threshold model has notable implications for applications such as viral marketing, political campaigning, and information dissemination. By capturing reinforcement phenomena like echo chambers and social amplification, PT can improve predictions of influence spread and help organizations select influential seed nodes more effectively. Consequently, it can inform policymaking for combating misinformation and optimizing public awareness campaigns, offering decision-makers a clearer understanding of how local network pressures shape overall influence dynamics.

\subsection{Future Work}
Future work includes PT‑specific approximation analyses. For non‑monotone PT objectives, the $1/e$ approximation factor provides a natural baseline. An important direction is to characterize regimes with limited non‑monotonicity, enabling improved, data‑dependent guarantees and recovery of the classical $(1-1/e)$ limit as $\alpha \to 0$. In addition, validating the PT model on large‑scale, time‑stamped diffusion datasets and estimation of the pressure parameter $\alpha$ from empirical traces are also important directions. Such empirical investigations can inform theory, algorithm design, and the regimes where pressure-driven effects dominate.

\subsection{Ethical Considerations}
\textcolor{black}{While the PT model provides a framework for understanding and optimizing information diffusion, such methods may also be applied to influence operations or targeted persuasion. We recognize the dual-use nature of this research and emphasize transparency, accountability, and fairness. When used in public communication or policy contexts, amplification and seed-selection mechanisms should be audited to prevent manipulative or discriminatory outcomes. All data used in this study are public and anonymized; no personally identifiable information was processed. }

\section*{Acknowledgments} 
We thank Camille Woods for valuable discussions on the PT model's theoretical properties.

\bibliography{aaai2026v2}
\end{document}